\documentclass[twocolumn]{aastex62}

\usepackage{amsmath}
\usepackage{amssymb}
\usepackage{textcomp,gensymb}
\usepackage{bm}
\usepackage{lineno}
\usepackage{float}
\usepackage{here}
\usepackage{graphicx}

\graphicspath{{./}{figures/}}

\submitjournal{ApJ}

\shorttitle{Theory of Stochastic Shock Drift Acceleration}
\shortauthors{Katou\& Amano}

\begin{document}

\title{Theory of Stochastic Shock Drift Acceleration for Electrons in the Shock Transition Region}

\newcommand{\utokyo}{
Department of Earth and Planetary Science, University of Tokyo, 7-3-1, Hongo, Bunkyo-ku, Tokyo, 113-0033, Japan
}
\correspondingauthor{Takuma Katou}
\email{t.kato@eps.s.u-tokyo.ac.jp}

\author{Takuma Katou}
\affil{\utokyo}

\author{Takanobu Amano}
\affil{\utokyo}



\begin{abstract}
We propose a novel electron acceleration mechanism, which we call stochastic shock drift acceleration (SSDA), that extends the standard shock drift acceleration (SDA) for low-energy electrons at a quasi-perpendicular shock to include the effect of stochastic pitch-angle scattering. We demonstrate that the steady-state energy spectrum of electrons accelerated within the shock transition region becomes a power-law in the limit of strong scattering. The spectral index is independent of the pitch-angle scattering coefficient. On the other hand, the maximum energy attainable through the mechanism scales linearly with the pitch-angle scattering coefficient. These results have been confirmed by Monte-Carlo simulations that include finite pitch-angle anisotropy. We find that the theory can reasonably well explain in-situ observations of quasi-perpendicular Earth's bow shock. Theoretical scaling law suggests that the maximum energy increases in proportion to the square of the shock speed, indicating that the thermal electrons may be accelerated up to mildly relativistic energies by the SSDA at quasi-perpendicular supernova remnant shocks. Therefore, the mechanism provides a plausible solution to the long-standing electron injection problem.
\end{abstract}

\keywords{acceleration of particles --- Earth --- shock waves}


\section{introduction}\label{sec:intro}
The acceleration of non-thermal particles has been a topic of great interest in space physics and astrophysics. The standard paradigm assumes that cosmic rays (CRs) of energies below $\sim 10^{15.5}$ eV are accelerated at supernova remnant (SNR) shocks in Galaxy via the so-called diffusive shock acceleration (DSA) mechanism \citep[e.g.,][]{1983RPPh...46..973D,1987PhR...154....1B}. The presence of relativistic electrons in young SNRs has been confirmed by radio and X-ray synchrotron emissions \citep{1995Natur.378..255K,2003ApJ...589..827B}. Observations of very-high-energy $\gamma$-rays may also be explained by the inverse-Compton photons emitted from the highly relativistic electrons \citep[e.g.,][]{2006A&A...449..223A}. Spectral modeling is often based on the DSA theory, which provides more or less adequate fits to the observed broadband spectra. Similarly, the theory has been tested against in-situ observations of energetic particles associated with shocks in the heliosphere \citep[e.g.,][]{1986JGR....9111917K}. Although protons are by far the major component of CRs detected on the Earth, the radiative signature of the hadronic component is much weaker than the leptonic counterpart. Therefore, understanding of the non-thermal emission from relativistic electrons is essential to probe physical parameters at the remote acceleration sites.

It is well known that the DSA is an efficient mechanism for accelerating sufficiently high energy particles. Therefore, in general, the mechanism requires a pre-accelerated seed population; the threshold energy depends heavily on the macroscopic shock parameters and the particle species. The DSA assumes that the particles are scattered by electromagnetic turbulence around the shock front so that the particle transport is described by diffusion relative to the background plasma flow. Therefore, one may presume that the condition for efficient pitch-angle scattering determines the threshold energy for the injection.

While pitch-angle scattering via the cyclotron resonance with magnetohydrodynamics (MHD) turbulence may occur relatively easily for protons, low energy electrons cannot satisfy the resonance condition because of their small gyroradii. As we will see later in Section~\ref{sec:discussion}, it is easy to confirm that mildly relativistic energy is needed for electrons resonantly interacting with MHD waves in parameters typical of interplanetary or interstellar media. This threshold energy is much larger than the expected temperatures in the downstream of, e.g., SNR shocks. Therefore, unless there exists a seed population with energies orders of magnitude higher than the thermal energy, efficient particle acceleration will not take place. This is the well-known electron injection problem that has been the subject of substantial debates over the decades.

One may roughly classify possible solutions to the electron injection issue proposed in literature into two categories. The first is to consider a particle acceleration mechanism which directly energizes the thermal particles into the relativistic energy range, so that the standard DSA may operate subsequently. Typically, large-amplitude plasma waves generated within the shock transition layer are invoked as the agent for an efficient particle acceleration process  \citep{1988ApJ...329L..29C,1995JPlPh..54...59G,2001PhRvL..87y5002M,2002ApJ...572..880H,2009ApJ...690..244A,2011ApJ...733...63R,2015Sci...347..974M}. The second possibility is to take into account the effect of high-frequency whistler waves as the scattering agent. Given sufficiently strong wave power in the frequency range that can resonantly scatter low-energy electrons, they may be accelerated by the DSA process. The problem is how to generate and maintain the power in the whistler wave frequency range, in particular, in the upstream of the shock \citep{1992ApJ...401...73L,1996MNRAS.278.1018L,2010PhRvL.104r1102A}. Unless there exists an efficient wave generation mechanism, waves at such high frequency will suffer strong cyclotron damping. Although substantial effort has been devoted to searching for the resolution, the problem yet remains controversial in both of the above scenarios. As we will see, the mechanism that we propose in this paper may be considered as a hybrid of these scenarios. In other words, we consider a direct particle acceleration mechanism in the shock transition layer in which high-frequency whistler waves play an important role.

Observations of energetic electrons in the heliosphere may give us a hint to resolve the issue. First of all, it is known that energetic electron flux enhancements associated with shock crossings are rare especially at interplanetary shocks \citep{2003AIPC..679..640L,2016A&A...588A..17D}. This is in clear contrast to protons and other heavy ions, for which the so-called gradual event, a signature expected from the DSA theory, is very common \citep{1980JGR....85.4602S,1985JGR....90....1T}. From the theoretical point of view, it is natural that the electron acceleration by the DSA is inefficient as there is no way to scatter low-energy electrons. Occasionally, however, substantial flux increases have been detected in the close vicinity of the shock, and such events are often called shock-spike events \citep[e.g.,][]{1989JGR....9410011G}. A statistical analysis of Earth's bow shock crossings showed that such signatures are seen almost exclusively at quasi-perpendicular shocks with $\theta_{Bn} \gtrsim 45 \degr$, where $\theta_{Bn}$ is the angle between the upstream magnetic field direction and the shock normal \citep{2006GeoRL..3324104O}. More specifically, \cite{2006GeoRL..3324104O} observationally found that there exists a critical Alfven Mach number beyond which the acceleration becomes efficient. The critical Mach number appears to be consistent with the one theoretically proposed by \cite{2010PhRvL.104r1102A}, which may be written as $M_{\rm A}^{\rm crit} \approx \cos \theta_{Bn} / 2 \sqrt{\beta_e m_i/m_e}$ where $\beta_e$, $m_i$, $m_e$ are respectively the electron plasma beta, proton mass, and electron mass. Recently, in-situ measurements made by the Cassini spacecraft at Saturn's bow shock extended the parameter space and confirmed that the observations are consistent with this idea even at higher Mach number quasi-parallel shocks \citep{2013NatPh...9..164M,2017ApJ...843..147M}. A typical energy range of energetic electrons seen at shock waves in the heliosphere is $\sim 1{\rm-}100$ keV but can reach nearly $\sim 1$ MeV at the highest Mach number Saturn's bow shock that has ever been observed in-situ \citep{2013NatPh...9..164M}. We emphasize that the typical energy range of energetic electrons measured in-situ at planetary bow shocks is indeed the energy range in between the thermal and threshold energy for the injection. Therefore, understanding the mechanism of particle acceleration in this regime will be crucial to resolve the issue of electron injection.

Conventionally, the shock drift acceleration (SDA) has been thought to play a role in generating energetic electrons in the upstream of Earth's bow shock \citep{1984JGR....89.8857W,1984AnGeo...2..449L}. The energetic electrons are typically streaming along the local magnetic field away from the bow shock. The flux is the most intense at those field lines connected to near the point of tangency between the magnetic field and the curved bow shock, suggesting that the acceleration efficiency is a sensitive function of $\theta_{Bn}$. As we explain in detail in section \ref{sec:adiabatic}, the energy gain via the SDA process is proportional to $1/\cos^2 \theta_{Bn}$, and the accelerated electrons are reflected back into upstream, forming a field-aligned beam. Although these properties qualitatively agree well with the observed characteristics, quantitative inconsistencies between the theory and observations yet remain. As pointed out by \cite{2001JGR...106.1859V}, the failure of the SDA model indicates that the assumption of adiabaticity needs to be revisited. Indeed, with the recent Magnetospheric Multiscale (MMS) spacecraft measurement of the bow shock, \cite{2017ApJ...842L..11O} clearly identified wave-particle interactions between suprathermal electrons and high-frequency whistler waves via the cyclotron resonance. Since the pitch-angle scattering by the wave-particle interaction breaks the conservation of the first adiabatic invariant, the theory must be modified to take into account such a non-adiabatic effect for more accurate modeling.

A yet another hint was provided by a recent fully self-consistent three-dimensional (3D) particle-in-cell (PIC) simulation of a high Mach number quasi-perpendicular shock by \cite{PhysRevLett.119.105101}. They found that non-thermal electrons were produced via the shock-surfing acceleration \citep[SSA;][]{2001PhRvL..87y5002M,2002ApJ...572..880H} followed by the SDA. The particles accelerated initially by the SSA had larger gyroradii comparable to the typical wavelength of large-amplitude Weibel turbulence generated by the reflected ions. Such particles were thus strongly scattered by the magnetic fluctuations within the shock transition region. The authors argued that the scattering enhanced the efficiency of the SDA because the particles were confined within the shock transition region longer than expected from the adiabatic SDA, although the energy gain mechanism itself was the same as the standard SDA. This particle acceleration process, which we call the stochastic shock drift acceleration (SSDA), is the main subject of this paper.

This paper presents a theoretical model of the SSDA, which is based on the standard SDA but takes into account the effect of pitch-angle scattering. We assume that phenomenological pitch-angle scattering occurs in such a way to conserve the kinetic energy in the plasma rest frame so that the energy gain itself comes entirely from the SDA. In contrast to the standard SDA, the spatially-averaged, steady-state energy spectrum becomes a power-law in the limit of strong scattering where pitch-angle anisotropy is negligible. We show that the spectral index does not depend on the pitch-angle scattering coefficient and is determined by a magnetic field gradient scale length. On the other hand, the maximum energy that may be obtained through the process is proportional to the efficiency of pitch-angle scattering. These theoretical predictions are confirmed by comparison with Monte-Carlo simulations that take into account finite pitch-angle anisotropy. We also discuss the application of the theoretical model to in-situ observations of Earth's bow shock.

This paper is organized as follows. In Section \ref{sec:theory}, the theoretical model is described. First, we briefly review the standard theory of the SDA, and then the effect of pitch-angle scattering is introduced. A box model is used for analytical tractability to investigate the spatially-averaged spectrum in the steady state. Monte-Carlo simulation results are presented in Section \ref{sec:simulation} to confirm the theoretical analysis. Section \ref{sec:discussion} presents a comparison between the theory and in-situ observations at Earth's bow shock. Finally, a summary is given in Section \ref{sec:summary}.


\section{Theory} \label{sec:theory}

\subsection{Adiabatic Shock Drift Acceleration} \label{sec:adiabatic}
In this subsection, we briefly review the standard theory of the SDA for low-energy electrons \citep{1984JGR....89.8857W,1984AnGeo...2..449L}. Let us consider the interaction of an electron with a magnetic field compression at a plane shock. It is well known that the thickness of the collisionless shock transition layer is roughly given by $u_{0}/\Omega_{\rm ci}$ calculated with the upstream flow speed $u_{0}$ measured in the normal incidence frame (NIF) and the ion cyclotron frequency $\Omega_{\rm ci}$ defined also with the upstream magnetic field strength \citep[e.g.,][]{1982JGR....87.5081L}. Here the NIF is defined as the reference frame at which the shock is at rest and the upstream flow velocity is parallel to the shock normal direction. For thermal and supra-thermal electrons with gyroradii much smaller than the shock thickness, one may assume that the adiabatic theory reasonably approximates the electron dynamics because the shock as seen from such electrons is no longer a discontinuity but a smooth magnetic field gradient. That is, the first adiabatic invariant $M = m_e v_{\perp}^2/ 2 B$ (defined with the perpendicular velocity $v_{\perp}$, and the magnetic field strength $B$) is constant during the interaction with the shock.

Now we consider the adiabatic interaction in the so-called de Hoffmann-Teller frame (HTF) where the motional electric field vanishes ($\bm{E} = - \bm{u} \times \bm{B} = 0$) both in the upstream and downstream. This substantially simplifies the analysis because the particle kinetic energy $\epsilon= m_e (v_{\parallel}^2 + v_{\perp}^2) / 2$ is also a conserved quantity in this frame. An electron traveling from the upstream toward the shock experiences a magnetic field gradient at the shock. If the pitch-angle (also defined in the HTF) is greater than the loss-cone angle $\theta_{\rm c}$ determined by the magnetic field compression at the shock, such electrons are reflected back upstream due to the magnetic mirror force. Although the electron energy is conserved in the HTF, the same process as viewed in a different frame makes a finite energy gain. More specifically, the gain in velocity in the upstream rest frame may be written as:
\begin{align} \label{csda}
 \Delta v = \frac{2 u_{\rm 0}}{\cos \theta_{B_{\rm n}}} = 2 u_{\rm sh},
\end{align}
where $u_{\rm sh} = u_{\rm 0}/\cos \theta_{B_{\rm n}}$ is the upstream plasma flow speed measured in the HTF. A substantial energy gain is thus possible at nearly perpendicular shocks ($\cos \theta_{B_{\rm n}} \ll 1$), as long as the shock is subluminal. On the other hand, for those particles transmitted to the downstream, the conservation of the first adiabatic invariant determines the energy gain. Namely, an energy gain only by a factor of four (at a strong shock) is the maximum that may be obtained through the process for the transmitted particles.

For the analysis presented in the next subsection, it is instructive to discuss the physical mechanism of the energy gain as seen in the NIF. \citet{1989JGR....9415367K} showed that the energy gain in the NIF is understood as that obtained by the gradient-$B$ drift in the direction anti-parallel to the motional electric field $-\bm{u} \times \bm{B}$. It is easy to check that the gradient-$B$ drift velocity given by
\begin{align}
 \bm{v}_{\nabla B} = \frac{M}{q} \bm{b} \times \bm{\nabla} \log B
\end{align}
(where $q$ is the particle charge and $\bm{b} = \bm{B}/B$ is the magnetic field unit vector) at a fast-mode shock is always in the direction such that the particle gains energy. Assuming that the magnetic field gradient is approximately constant during the interaction, and the shock structure is nearly stationary (where $-\bm{u} \times \bm{B}$ is also constant), we can easily understand that the final energy gain of the particle is proportional to the interaction time. Therefore, the distinction between the reflected and transmitted electrons may be understood by the difference in the interaction time. For the transmitted electrons, the dominant $\bm{E} \times \bm{B}$ drift quickly transports particles to the downstream. On the other hand, the mirror force directed toward upstream competes with the convection, which makes the interaction time for the reflected particles much longer and the energy gain becomes much more efficient.

Note that, in the above analysis, we have ignored the effect of the cross-shock electrostatic potential. This may be justified as long as the final energy gain of the reflected particles is concerned. However, it changes an effective loss cone angle that determines whether the electrons are reflected or transmitted. Nevertheless, it is known that this effect is not significant at energies much larger than the potential. Therefore, in this paper, we consider only those electrons for which the cross-shock potential effect is negligible.

As we mentioned, the adiabatic SDA qualitatively well explains the observations near Earth's bow shock \citep[e.g.][]{1979GeoRL...6..401A,1981JGR....86.4343P}. However, the observed fluxes, spectra, and anisotropies of energetic electrons cannot be explained quantitatively by the simple theory alone. \citet{2001JGR...106.1859V} suggested that violation of the first adiabatic invariant via pitch-angle scatterings may enhance the rate of energy gain through the process. The inclusion of probabilistic nature makes it possible to confine a small fraction of electrons within the acceleration region more efficiently. Since longer interaction time increases the final energy gain of particles, more efficient particle confinement will result in more efficient particle acceleration. This motivates our analysis of the SSDA theory presented in the following subsections.

\subsection{Effect of Pitch-angle Scattering} \label{sec:model}
Since the pitch-angle scattering introduces stochasticity into the energization process, the problem is the best described by using a transport equation for the distribution function. We start with the standard focused transport equation with a phenomenological pitch-angle scattering term \citep{1975MNRAS.172..557S,1997JGR...102.4719I}:
\begin{align} \label{original_transport}
\frac{\partial f}{\partial t}
&+(\bm{u}+v\mu\bm{b})\cdot\nabla f \notag\\
&+
\left[
\frac{1-3\mu^2}{2}\bm{b}\bm{b}:\nabla\bm{u}-\frac{1-\mu^2}{2}\nabla\cdot\bm{u}\right.\notag\\
&\left.-\frac{\mu\bm{b}}{v}\cdot\left(\frac{\partial \bm{u}}{\partial t}+\bm{u}\cdot\nabla\bm{u}\right)
\right]
v\frac{\partial f}{\partial v}\notag\\
&+\frac{1-\mu^2}{2}
\left[
v\nabla\cdot\bm{b}+\mu\nabla\cdot\bm{u}-3\mu\bm{b}\bm{b}:\nabla\bm{u}\right.\notag\\
&\left.-\frac{2\bm{b}}{v}\cdot\left(\frac{\partial \bm{u}}{\partial t}+\bm{u}\cdot\nabla\bm{u}\right)
\right]
\frac{\partial f}{\partial \mu}\notag\\
&=\frac{\partial}{\partial \mu}
\left[
(1-\mu^2)D_{\mu\mu}\frac{\partial f}{\partial\mu}
\right]+Q,
\end{align}
where $f(\bm{x},v,\mu,t)$ is the gyrotropic part of the phase-space density and $Q$ is the source term. The spatial coordinate $\bm{x}$ is measured in the inertial frame, whereas the magnitude of particle velocity $v$ and pitch angle cosine $\mu$ are both measured in the local plasma rest frame. The first term on the right-hand side describes diffusion in pitch angle with the coefficient $D_{\mu\mu}$. Note that $D_{\mu\mu}^{-1}$ represents the characteristic time scale with which pitch-angle anisotropy relaxes toward isotropy.

Now, let us assume the presence of the HTF for which the plasma flow velocity is everywhere parallel to the local magnetic field $\bm{u}=u_{\parallel}\bm{b}$ (where $u_{\parallel}$ represents the flow speed parallel to the magnetic field). It is easy to confirm that the inertial term proportional to $\left(\partial/\partial t + \bm{u} \cdot \nabla \right)\bm{u}$ in the focused transport equation is negligible for supra-thermal electrons with $v \gg u_{sh}$ in a time independent flow ($\partial\bm{u}/\partial t=0$). This simplifies the transport equation in the following form:
\begin{align}\label{middle_transport}
\frac{\partial f}{\partial t}&+(v\mu+u_{\parallel})\frac{\partial f}{\partial s}\notag\\
&+\left[\frac{1-\mu^2}{2}\frac{\partial \log B}{\partial s}u_{\parallel}-\mu^2\frac{\partial u_{\parallel}}{\partial s}\right]v\frac{\partial f}{\partial v}\notag\\
&-\frac{1-\mu^2}{2}\left[(u_{\parallel}\mu+v)\frac{\partial\log B}{\partial s}+2\mu\frac{\partial u_{\parallel}}{\partial s}\right]\frac{\partial f}{\partial \mu}\notag\\
&=\frac{\partial}{\partial \mu}
\left[
(1-\mu^2)D_{\mu\mu}\frac{\partial f}{\partial \mu}
\right]+Q,
\end{align}
where $s$ denotes the spatial coordinate along the magnetic field line.

It is important to note that Equation (\ref{middle_transport}) may reproduce the standard DSA in an appropriate limit. Namely, in the limit of isotropic pitch-angle distribution at a parallel shock (where $B$ is constant), the transport equation reduces to the standard diffusion-convection equation for energetic particles. In this case, a negative flow divergence $\partial u_{\parallel}/\partial s < 0$ provides the only source of particle energy gain. At an oblique shock, the rate of energy gain through the DSA increases although the spectral index does not change \citep{1987ApJ...313..842J}. In the HTF, the increase in the energy gain rate may be understood as the contribution from a finite magnetic field gradient $\partial \log B/\partial s > 0$.

The standard DSA assumes that the shock is discontinuous or, in other words, the shock thickness is much smaller than the typical mean free path of the accelerated particles. In contrast, our purpose is to take into account the effect of pitch-angle scattering while low-energy electrons are interacting with the shock structure itself. The flow velocity and the magnetic field are thus considered to be smooth functions of $s$.

In this paper, we focus our discussion primarily on quasi-perpendicular shocks. This is motivated by the bow shock observations where quasi-perpendicular shocks are usually more efficient in accelerating electrons than quasi-parallel shocks. As is detailed in Appendix \ref{sec:appendix}, the ratio between velocity and magnetic field gradient scale lengths may be estimated as follows:
\begin{align} \label{estimation}
&\left |\left(\frac{\partial \log{u_{\parallel}}}{\partial s}\right)\left(\frac{\partial \log B}{\partial s}\right)^{-1}\right|\notag\\
&\approx\frac{1}{2\log{r}}\left(1-\frac{1}{r^2}\right)\cos^2{\theta_{B_{\rm n}}}+O\left(\cos^4{\theta_{B_{\rm n}}}\right).
\end{align}
This indicates that the contribution from the velocity divergence may be ignored at a quasi-perpendicular shock with $\cos \theta_{B_{\rm n}} \ll 1$. Therefore, we assume $u_{\parallel} = u_{\rm sh}$ to be constant in the entire shock transition region in the following discussion.

\begin{figure}[!ht]
\centering
\includegraphics[width=0.4\textwidth]{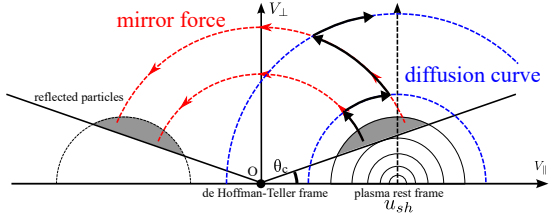}
\caption{Schematic illustration for the particle trajectory accelerated by the SSDA in velocity space. The velocity is defined in the HTF. The red curves indicate
the particle trajectory expected from the standard SDA (or adiabatic mirror reflection). The pitch-angle scattering induces diffusion of particles
along the diffusion curve shown in blue. The thick arrows represent a schematic trajectory of an electron accelerated by the SSDA. The
velocity distribution function of the upstream population is shown in contours. In the absence of scattering, the mirror reflection occurs only for
those particles outside the loss cone (pitch-angle larger than the loss cone angle $\theta_{\rm c}$) shown in the gray-hatched area.
\label{fig:ssda}}
\end{figure}

Under these assumptions, we finally obtain the following electron transport equation:
\begin{align}\label{final_transport}
\frac{\partial f}{\partial t}
&+(v\mu+u_{\rm sh})\frac{\partial f}{\partial s}
+\frac{1-\mu^2}{2}\frac{\partial\log B}{\partial s}u_{\rm sh}v\frac{\partial f}{\partial v}\notag\\
&-\frac{1-\mu^2}{2}\frac{\partial\log B}{\partial s}(u_{\rm sh}\mu+v)\frac{\partial f}{\partial \mu}\notag\\
&=\frac{\partial}{\partial \mu}\left[(1-\mu^2)D_{\mu\mu}\frac{\partial f}{\partial \mu}\right]
+ Q.
\end{align}
It is easy to confirm that the velocity and pitch-angle changes in the absence of pitch-angle scattering ($D_{\mu\mu} = 0$) gives the simple adiabatic mirror reflection, in which the particle trajectory follows a circle in $v_{\parallel} {\rm -} v_{\perp}$ plane with its center at the origin in the HTF. (The particle trajectory for the mirror reflection is schematically illustrated in Figure \ref{fig:ssda} as the red dashed curve.) The model thus reduces to the standard SDA in this limit. 

The crucial point is that the pitch-angle scattering occurs in such a way to conserve the particle energy in the plasma rest frame (scattering induces diffusion of particles along the blue curves in Figure \ref{fig:ssda}), whereas the mirror reflection (or SDA) conserves the energy in the HTF. Therefore, the energy gain comes from the velocity difference between the two moving {\it magnetic walls} that are approaching with each other. This is similar to the standard DSA where the scatterings both in the upstream and downstream play the role of the moving walls. However, the rate of energy gain via the SSDA is much more efficient. This is because the accelerated electrons are always trapped in the shock transition region where they experience the constant magnetic mirror force. In contrast, the energy gain rate via the DSA is much slower because particles must traverse the shock front back and forth. The pitch-angle scattering tends to confine the accelerated electrons, which may otherwise quickly escape from the acceleration region.  As we will see below, the increase in the interaction time with the shock via sufficiently strong pitch-angle scatterings leads to the formation of a power-law spectrum. The overall acceleration efficiency is, therefore, substantially increased.

Note that we have ignored the energy gain by the 2nd order Fermi acceleration process by assuming the elastic scattering in the plasma rest frame. This is because the phase velocity of whistler waves both in the typical interstellar and interplanetary media ($\sim 1000$ km/s) is much smaller than the velocities of accelerated electrons.
\subsection{A Box Model} \label{sec:box}

In principle, one may numerically solve Equation (\ref{final_transport}) to obtain the electron spectrum including anisotropy as a function of the field-aligned coordinate $s$ in the shock transition region. However, our primary purpose is to understand an overall picture of the SSDA mechanism which does not necessarily need such details.

We employ a box model to simplify the problem to an analytically tractable form. Note that a similar approach may also be used for the standard DSA model \citep[e.g.][]{1999A&A...347..370D}, indicating that this simplification keeps the essential physics of the particle acceleration. In short, we consider a spectrum averaged over the shock transition region. We will also assume that the scattering is strong enough so that the particle pitch-angle distribution is nearly isotropic.  In the following discussion, we assume $D_{\mu\mu}$ is constant for simplicity. The validity of this assumption will be discussed later in Section \ref{sec:discussion}.

Since we are not interested in the internal structure of the shock transition region, we introduce the spatial scale length $L$ representing an averaged magnetic field gradient scale length:
\begin{align}
 \frac{1}{L} \equiv \frac{\partial\log B}{\partial s},
\end{align}
which is constant over the region of interest.

It is convenient to rewrite Equation (\ref{final_transport}) into the conservation form:
\begin{align}\label{flux_form}
\frac{\partial N}{\partial t}
&+\frac{\partial}{\partial s}\left[(v\mu+u_{\rm sh})N\right]+\frac{\partial}{\partial v}\left[\frac{1-\mu^2}{2L}u_{\rm sh}vN\right]\notag\\
&-\frac{\partial}{\partial\mu}\left[\frac{1-\mu^2}{2L}\left(v+u_{\rm sh}\mu\right)N\right]
\notag \\
&=\frac{\partial}{\partial \mu}\left[(1-\mu^2)D_{\mu\mu}\frac{\partial N}{\partial \mu}\right]+Q,
\end{align}
where $N(s, v,\mu)= 2 \pi f(v,\mu) v^2 \exp(-s/L)$. Note that $N(v,\mu) dv d\mu$ represents the number of electrons in the phase-space volume $dv d\mu$ integrated over the cross-sectional area $A = \exp(-s/L)$ of the magnetic flux tube. We now define the spatial range of the shock transition region as $0 < s < L_{\rm sh}$ with $s = 0$ and $s = L_{\rm sh}$ being the upstream and downstream edges, respectively. We adopt the approximation of the length scale $L_{\rm sh}$ of the shock transition region along the magnetic field line:
\begin{align}\label{shocksize}
 L_{\rm sh} \approx \frac{u_{0}}{\Omega_{\rm ci}\cos{\theta_{B_{\rm n}}}}=\frac{u_{\rm sh}}{\Omega_{\rm ci}}.
\end{align}

By taking average over $s$ in the range $0 \le s \le L_{\rm sh}$, we obtain
\begin{align} \label{montecarlo}
 \frac{\partial \bar{N}}{\partial t}
&+r_{\rm esc} \bar{N}
+\frac{\partial}{\partial v}\left[\frac{1-\mu^2}{2L}u_{\rm sh}v \bar{N}\right]\notag\\
&-\frac{\partial}{\partial\mu}\left[\frac{1-\mu^2}{2L}\left(v+u_{\rm sh}\mu\right) \bar{N}\right]\notag\\
&=\frac{\partial}{\partial \mu}\left[(1-\mu^2)D_{\mu\mu}\frac{\partial \bar{N}}{\partial \mu}\right] + \bar{Q},
\end{align}
where the spatially-averaged spectrum $\bar{N} (v,\mu)$ is defined as follows
\begin{align}
 \bar{N} (v,\mu) \equiv
 \frac{1}{L_{\rm sh}} \int_0^{L_{\rm sh}} ds \, N(s, v, \mu).
\end{align}
$\bar{Q}$ is also defined similarly by taking spatial average of the source term $Q$. The second term on the left-hand side represents the particle escape from the boundaries with the {\it escape rate} $r_{\rm esc} (v,\mu)$ defined by
\begin{align}
r_{\rm esc} \bar{N} \equiv
\frac{1}{L_{\rm sh}} \int_{0}^{L_{\rm sh}} ds
\frac{\partial}{\partial s} \left[ (v\mu+u_{\rm sh}) N(s,v,\mu) \right].
\end{align}

Since it is difficult to estimate the escape rate $r_{\rm esc}$ for a distribution function of arbitrary anisotropy, we consider the limit of strong pitch-angle scattering. In other words, we consider the isotropic part of the pitch-angle distribution and assume that the anisotropy is negligible. The isotropic part of the spectrum may be obtained by averaging over pitch angle $\mu$:
\begin{align}
 \bar{N}_0 (v) \equiv
\frac{1}{2} \int_{-1}^{+1} d\mu \, \bar{N} (v, \mu).
\end{align}
The transport equation for $\bar{N}_0 (v)$ is obtained by further taking average of Equation~(\ref{montecarlo}) over $\mu$:
\begin{align}\label{box1}
\frac{\partial \bar{N}_0}{\partial t}
+\frac{\partial}{\partial v} \left( \frac{u_{\rm sh} v}{3 L} \bar{N}_0 \right)
= \bar{Q} - r_{\rm esc,0} \bar{N}_0,
\end{align}
where we have assumed an isotropic injection $\bar{Q} (v)$. $r_{\rm esc,0}$ is the escape rate for the isotropic part of the distribution function defined as follows
\begin{align}
r_{\rm esc,0} \bar{N}_0
& \equiv
\frac{1}{2} \int_{-1}^{+1} d\mu \frac{1}{L_{\rm sh}} \int_{0}^{L_{\rm sh}} ds
\frac{\partial}{\partial s} \left[ (v\mu+u_{\rm sh}) N(s,v,\mu) \right]\notag\\
&=
\left.
\frac{1}{2 L_{\rm sh}} \int_{-1}^{+1} d\mu [(v\mu+u_{\rm sh}) N(s,v,\mu)]
\right|_{s=L_{\rm sh}} \notag\\
&-\left.
\frac{1}{2 L_{\rm sh}} \int_{-1}^{+1} d\mu [(v\mu+u_{sh}) N(s,v,\mu)]
\right|_{s=0} \notag\\
& \equiv
\left( r_{\rm esc,0}^{\rm down} + r_{\rm esc,0}^{\rm up} \right) \bar{N}_0 (v).
\end{align}
$r_{\rm esc,0}^{\rm up}, r_{\rm esc,0}^{\rm down}$ describe the rate of escape at the upstream and downstream boundaries of the shock transition region, respectively. These must be determined by specifying appropriate boundary conditions.

We first consider the escape toward the downstream $r_{\rm esc,0}^{\rm down}$. For this, we assume that the spectrum is continuous across the downstream boundary which implies that the electron pitch-angle distribution in the downstream is also isotropic. In this case, we may adopt the approximation $N(L_{\rm sh}, v, \mu) \approx \bar{N}_0 (v)$, with which we obtain
\begin{align} \label{descape}
 r_{\rm esc,0}^{\rm down}
 \approx
 \frac{1}{2 L_{\rm sh} \bar{N}_0}
 \int_{-1}^{+1} d \mu \left[ (v \mu + u_{\rm sh}) N(L_{\rm sh}, v, \mu) \right]
 = \frac{u_{\rm sh}}{L_{\rm sh}}
\end{align}

To estimate the escape rate toward the upstream $r_{\rm esc,0}^{\rm up}$, we need to consider the spatial distribution of electrons within the shock transition region. The strong scattering limit implies that the mean free path of electrons $\lambda_{\rm mfp}=v/2D_{\mu\mu}$ is much smaller than the system size $L_{\rm sh}$. Therefore, the spatial transport of the particles may be described by diffusion with the coefficient $\kappa$ given by \citep{1972ApJ...172..319J,1975MNRAS.172..557S}
\begin{align} \label{diffusion_app}
 \kappa=\frac{v^2}{6D_{\mu\mu}}.
\end{align}

Considering the balance between the convection (toward downstream) and the diffusion, we expect that the particle intensity increases exponentially toward downstream $\propto \exp(s/l_{\rm diff})$, where the spatial scale length of diffusion $l_{\rm diff}$ given by
\begin{align}\label{diffusion_length}
 l_{\rm diff} = \frac{\kappa}{u_{\rm sh}}=\frac{v^2}{6D_{\mu\mu}u_{\rm sh}}.
\end{align}
This indicates that the escape at the upstream boundary is negligible as long as the diffusion length is much smaller than the system size: $l_{\rm diff} \ll L_{\rm sh}$. However, the diffusion length increases as increasing the particle energy (for a constant $D_{\mu\mu}$), and at some point, the particles start to escape toward upstream. We approximate the distribution function at the boundary as $N(0, v, \mu) \approx \bar{N}_0 (v) \exp(-L_{\rm sh}/l_{\rm diff})$. Assuming that there is no incoming flux to the shock transition region from the upstream boundary, we obtain
\begin{align} \label{uescape}
 r_{\rm esc,0}^{\rm up}
 &\approx
 - \frac{1}{2 L_{\rm sh} \bar{N}_0}
 \int_{-1}^{-u_{\rm sh}/v} d \mu \left[ (v \mu + u_{\rm sh}) N(0, v, \mu) \right]
 \notag \\
 &= \frac{1}{4 L_{\rm sh}}
 \left(
 v - 2u_{\rm sh} \left( 1 - \frac{u_{\rm sh}}{2 v} \right)
 \right) \exp\left( -\frac{L_{\rm sh}}{l_{\rm diff}} \right).
\end{align}
In contrast to the escape toward downstream (Equation~(\ref{descape})), the upstream escape rate Equation~(\ref{uescape}) is energy dependent for $l_{\rm diff} \gtrsim L_{\rm sh}$.


\subsection{Steady-state Spectrum} \label{sec:steadystate}
We now analyze the steady-state spectrum. In the energy range where the escape toward upstream is negligible (i.e., $l_{\rm diff} \ll L_{\rm sh}$), the escape rate is determined only by the downstream escape $r_{\rm esc,0} \approx r_{\rm esc,0}^{\rm down}$. Assuming the spectrum of power-law form $\bar{N}_0 (v) \propto v^{-p}$ and considering the energy range far from the injection ($\bar{Q} \approx 0$), we obtain the spectral index:
\begin{align}\label{powerlaw}
 p=1+3\frac{L}{L_{\rm sh}}=1+3\left(L_{\rm sh}\frac{\partial\log{B}}{\partial s}\right)^{-1}.
\end{align}
We see that the spectral index $p$ is constant and is determined only by the scale length of the magnetic field gradient $L$. Since, by definition, it is a quantity on the order of $L_{\rm sh}$, we may roughly estimate the spectral index to be $p \sim 4$. This may be converted to the spectral index for the phase space density as a function of energy $f(\epsilon) \propto \epsilon^{-p/2 - 1} \sim \epsilon^{-3}$, which is close to the indices observed at Earth's bow shock \citep{1989JGR....9410011G,2006GeoRL..3324104O}. Note that we use the terms energy and velocity interchangeably in the following discussion as this will not cause any confusion.

The electron energy spectrum continues to follow the power-law as long as the upstream escape is negligible. However, if the diffusion length $l_{\rm diff}$ becomes comparable to the system size $L_{\rm sh}$, the accelerated electrons cannot be confined by the pitch-angle scattering and start to escape from the system. As we have already mentioned, the upstream escape rate $r_{\rm esc,0}^{\rm up}$ in this regime is clearly energy dependent, and the higher energy particles escape more efficiently than the lower energy. Therefore, this will break the power-law and introduce a cut-off in the spectrum. The cut-off energy $\epsilon_{\rm max}$ may be obtained by the condition $l_{\rm diff} \sim L_{\rm sh}$, yielding
\begin{align}\label{cutoff}
 \epsilon_{\rm max} \sim
 \epsilon_{\rm sh} \frac{m_{\rm i}}{m_{\rm e}} \frac{D_{\mu\mu}}{\Omega_{ce}}
 =
 \frac{1}{2}m_{\rm i} u_{\rm sh}^2\frac{D_{\mu\mu}}{\Omega_{\rm ce}}, 
\end{align}
where $\Omega_{\rm ce} = \Omega_{\rm ci}\,m_{\rm i}/m_{\rm e}$ is the electron cyclotron frequency. Note that if the maximum energy is normalized to $\epsilon_{\rm sh} = m_{\rm e} u_{\rm sh}^2 / 2$, it is determined by the single parameter $D_{\mu\mu}$.

In summary, the SSDA will produce a power-law spectrum with a maximum energy cut-off. The spectral index does not depend on the pitch-angle scattering coefficient $D_{\mu\mu}$ as long as the scattering is strong enough. This justifies our assumption of constant $D_{\mu\mu}$. On the other hand, the maximum energy depends linearly on $D_{\mu\mu}$. Therefore, the energy dependence of the pitch-angle scattering coefficient is indeed crucial for estimating the maximum energy, which will be discussed in Section \ref{sec:discussion}. It is important to mention that the dependence of the maximum energy on $\epsilon_{\rm sh}$ indicates that higher shock speeds and more oblique shocks will produce higher energy particles even for a fixed $D_{\mu\mu}$.

\section{Monte Carlo Simulation} \label{sec:simulation}
\subsection{Simulation Method and Setup}\label{sec:monte}
In the previous section, we have analyzed the spatially-averaged spectrum of the electrons under the assumption of isotropy. This assumption may, however, lose its validity at around the cut-off energy because the cut-off appears due to insufficient scattering. We thus solve Equation (\ref{montecarlo}) by using the Monte-Carlo method to take into account the effect of anisotropy. For the escape rate, however, we assume $r_{\rm esc} \approx r_{\rm esc,0}$. In general, the escape rate will deviate from the isotropic case if the anisotropy becomes non-negligible. A more rigorous treatment requires solving the transport equation including the spatial dependence, which is beyond the scope of the present paper.

To conduct Monte-Carlo simulations, we need to formulate the problem in terms of stochastic differential equations (SDEs). The equation of motion for an individual electron, which has both deterministic and stochastic terms, may be written as \citep[e.g.,][]{2012ApJ...745...62L}
\begin{align}
dv
&=\frac{1-\mu^2}{2L}u_{\rm sh}{v} \, dt, \label{dv}\\
d\mu
&=\left[-\frac{1-\mu^2}{2L}(u_{\rm sh}\mu+v)-2D_{\mu\mu}\mu\right] \, dt
+\sqrt{(1-\mu^2)D_{\mu\mu}} \, dW,\label{dmu}
\end{align}
where $dW$ denotes the Wiener process. In other words, it describes the Browninan motion which produces the Gaussian probability density function with mean $= 0$ and variance $= t$ at time $t$. We used the Milstein method \citep{gardiner2009stochastic} to integrate the SDEs. This method is one of the numerical It\^{o} integral of SDEs which has both strong and weak convergence to the first order in time step $\Delta t$. The escape from the system was implemented by removing electrons at each time step with the probability
\begin{align} \label{escape_rate}
 P_{\rm esc} = r_{\rm esc} \Delta t.
\end{align}

We used a particle splitting method to improve counting statistics at high energy \citep[e.g.,][]{1987ApJ...322..256K}. In the simulations presented in this paper, we split one particle into two identical particles with a half weight when the number of particles in the system becomes less than a half. Note that the results were not sensitive to the technical details of the particle splitting.

The simulations were initialized with $10^5$ particles that were isotropically injected into the system with an initial speed of $v_{\rm inj}/v_{\rm sh} = 1$. By integrating the SDEs for the particles, we obtain a numerical Green's function $G(t, v, \mu)$ at time $t$ for an impulsive injection at $t=0$. The distribution function resulting from a constant injection is then given by integrating the Green's function in time
\begin{align}
 \bar{N} (t, v, \mu) = \int_{0}^{t} G(\tau, v, \mu) d \tau.
\end{align}

The simulations were performed with a time step of $D_{\mu\mu}\Delta t = 5\times 10^{-4}$ until the solutions reached steady states. We checked that the results presented below did not depend on the time step and the number of particles used.

\subsection{Results}\label{sec:results}
\begin{figure}[t!]
\centering
\includegraphics[width=0.45\textwidth]{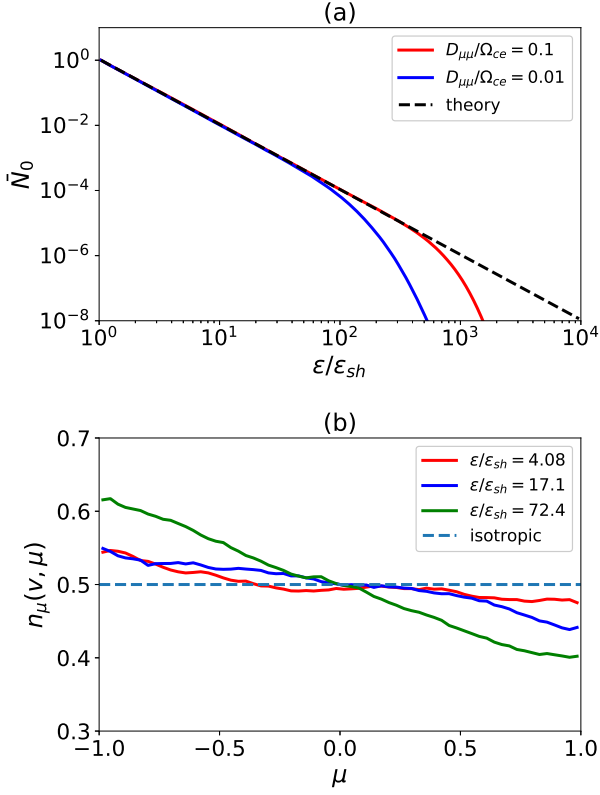}
\caption{(a) The energy spectra of electrons obtained with a fixed $L/L_{\rm sh}=1$ for $D_{\mu\mu}/\Omega_{\rm ce}=0.01, 0.1$. The dashed line represents a theoretical power-law for reference. (b) The normalized pitch-angle distribution for selected energies of $\epsilon/\epsilon_{sh}=4.08,17.1,72.4$ obtained with $L/L_{sh}=1$, $D_{\mu\mu}/\Omega_{\rm ce}=0.01$.}
\label{fig:spectrum}
\end{figure}

Figure~\ref{fig:spectrum}(a) shows the isotropic component of normalized electron energy spectra obtained with $L/L_{\rm sh}=1$ for two different pitch-angle scattering coefficients $D_{\mu\mu}/\Omega_{\rm ce}=0.01,0.1$. We normalized the distribution function at the injection energy $\epsilon/\epsilon_{\rm sh}=1$ to be unity. We observe a clear power-law spectrum with a cut-off at high energy in both cases. The power-law index agrees quite well with the theoretical index predicted by Equation (\ref{powerlaw}) shown in the dashed line for reference.  

Figure~\ref{fig:spectrum}(b) shows the normalized pitch-angle distribution $n_{\mu}(v,\mu)$ defined by
\begin{align}
n_{\mu}(v,\mu)=\frac{\bar{N}(t,v,\mu)}{{\displaystyle \int_{-1}^{1}\bar{N}(t,v,\mu)d\mu}}
\end{align}
at selected energies of $\epsilon/\epsilon_{\rm sh}=4.08,17.1,72.4$ for the run with $D_{\mu\mu}/\Omega_{\rm ce}=0.01$, $L/L_{\rm sh}=1$. The dashed line represents the isotropic distribution for reference. We see that the pitch-angle anisotropy is weak at low energies, but gradually increases as increasing the electron energy, especially around the maximum energy cut-off.
This is indeed an expected behavior. According to Equation (\ref{dmu}), the contribution of the SDA for the $\mu$ transport can be given by 
\begin{align}\label{mu_d}
d\mu_{\text{SDA}}=-\frac{1-\mu^2}{2L}\left(u_{\rm sh}\mu+v\right)dt\approx -\frac{1-\mu^2}{2L}vdt,
\end{align}
\begin{figure}[t!]
\centering
\includegraphics[width=0.45\textwidth]{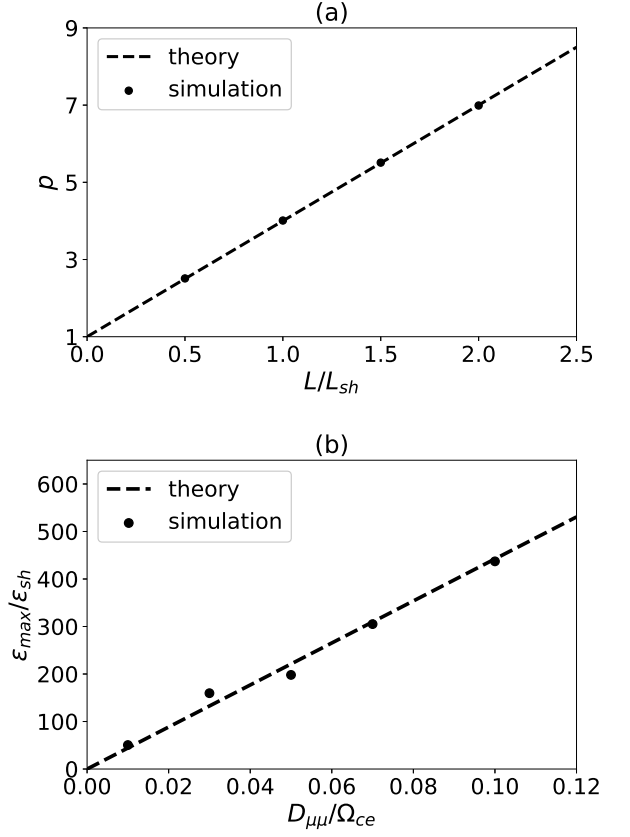}
\caption{(a) The dependence of the spectral index on $L/L_{\rm sh}$. The filled circles represent the simulation results obtained with $L/L_{\rm sh}=0.5,\,1,\,1.5,\,2.0$ for a fixed $D_{\mu\mu}/\Omega_{\rm ce}=0.01$. The dashed line shows the theoretical prediction Equation (\ref{powerlaw}) for reference. (b) The maximum cut-off energy as a function of $D_{\mu\mu}$. The simulation results obtained with $L/L_{sh}=1$, $D_{\mu\mu}/\Omega_{ce}=0.01,\,0.03,\,0.05,\,0.07,\,0.1$ are shown. The dashed line shows the result of fitting by a linear function predicted by the theory.}
\label{fig:maximum}
\end{figure}
for $v\gg u_{\rm sh}$. This explains the transport of particles toward the negative $\mu$ direction.
Nevertheless, this weak anisotropy has a negligible impact on the spectral index, which may be understood from the observed nearly linear anisotropy ($\propto -\mu$). Note that the spectral index is determined by the balance between the pitch-angle averaged acceleration and escape rates. The rate of escape toward downstream Equation (\ref{descape}) does not depend on the anisotropy. On the other hand, Equation (\ref{montecarlo}) indicates that the acceleration rate is an even function of $\mu$. Therefore, the observed nearly linear anisotropy ($\propto -\mu$) does not affect the pitch-angle averaged acceleration rate, so does not the power-law index.

We have performed parameter survey to check the consistency between the theory and simulations. Figure~\ref{fig:maximum}(a) shows the dependence of the spectral index $p$ on the magnetic field scale length $L/L_{\rm sh}$. The simulation results shown in filled circles determined by a power-law fit at low energy agree quite well with the theoretical prediction by Equation (\ref{powerlaw}). We have confirmed that the spectral index does not depend on other parameters as expected from the theory.

Figure~\ref{fig:maximum}(b) represents the maximum energy cut-off as a function of $D_{\mu\mu}$. We determined the cut-off energy as the energy at which the simulation result deviates from the power-law fit by 10\%. It is clear that the simulation results are consistent with the theoretical scaling ($\epsilon_{\rm max}\propto D_{\mu\mu}$) shown with the dashed line determined by a linear fit. We have confirmed that the cut-off energy depends only on the pitch-angle scattering coefficient, which is again consistent with the theory.

\section{Discussion} \label{sec:discussion}
As we have seen in the previous sections, the SSDA can produce a power-law spectrum of electrons accelerated within the shock transition region. The spectrum will have a maximum energy cut-off corresponding to the energy beyond which the assumed strength of pitch-angle scattering is no longer able to confine the particles in the acceleration region. So far, we have assumed that the pitch-angle scattering coefficient $D_{\mu\mu}$ is independent of particle energy and pitch angle. However, it is crucial to investigate the dependence for comparison with in-situ observations of Earth's bow shock.

In the following, we use the standard quasi-linear theory (QLT) and assume that the electrons are scattered by parallel propagating whistler waves via the cyclotron resonance. We note that the assumptions of QLT such as small-amplitude waves, random phase approximation, and the spatial homogeneity are not necessarily satisfied in the shock transition region. Also note that the whistler waves, in general, have oblique propagation angles with respect to the ambient magnetic field, which introduces Landau and higher-harmonic cyclotron resonances. Therefore, we think that the estimate based on QLT provides only an order of magnitude estimate of $D_{\mu\mu}$, which is nevertheless correct at least qualitatively.

The cyclotron resonance condition for an electron with a circularly polarized electromagnetic wave (with frequency $\omega$ and wavenumber $k$) is given by
\begin{align}\label{resonance}
 \omega-kv\mu=\frac{\Omega_{\rm ce}}{\gamma},
\end{align}
where $\gamma$ is the Lorentz factor. Notice that the positive (negative) frequency indicates the right-hand (left-hand) polarization. The dispersion relation for right-hand circularly-polarized electromagnetic waves propagating parallel to the ambient magnetic field is given by
\begin{align}\label{dwhistler}
\left(\frac{ck}{\omega}\right)^2=\frac{\omega_{\rm pe}^2}{(\Omega_{\rm ce}-\omega)\omega}-\frac{\omega_{\rm pi}^2}{(\Omega_{\rm ci}+\omega)\omega},
\end{align}
where $\omega_{\rm pe}$, $\omega_{\rm pi}$, and $c$ represent the electron and ion plasma frequencies, and the speed of light, respectively. The dispersion relation adopts a cold plasma approximation and describes only a low-frequency branch under the assumption $\omega / kc \ll 1$.

By combining Equations (\ref{resonance}) and (\ref{dwhistler}), we obtain the relation between the electron velocity and the wavenumber/frequency that satisfies the cyclotron resonance. If we fix a value of $V_{\rm A}/c$ (where $V_{\rm A}$ denotes the Alfven speed), this relation is uniquely determined. Figure \ref{fig:tresonance} shows the relation for the right-hand polarized wave $(\omega > 0)$ for three different values of $V_{\rm A}/c = 10^{-3}, 10^{-4}, 10^{-5}$. We used a fixed $|\mu| = 1$ and a realistic proton-to-electron mass ratio $m_{\rm i}/m_{\rm e} = 1836$. Note that the wavenumber is normalized to the ion inertial length $\lambda_{\rm i} = V_{\rm A}/\Omega_{\rm ci}$.

\begin{figure}[!htb]
\centering
\includegraphics[width=0.45\textwidth]{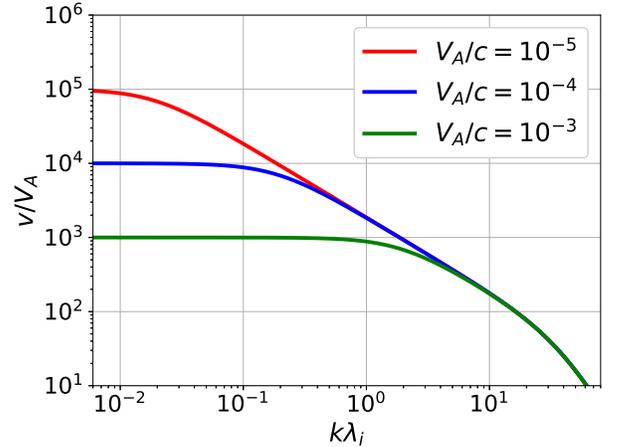}
\caption{The relationship between the resonance velocity and wavenumber of right-hand circularly polarized electromagnetic waves for three different
values of $V_{\rm A}/c = 10^{-5}, 10^{-4}, 10^{-3}$. The real proton to electron mass ratio $m_{\rm i}/m_{\rm e}$ is used.}
\end{figure}
\label{fig:tresonance}

It is easy to see that the resonance with the wave in the MHD regime $k \lambda_{\rm i} \ll 1$ occurs at $v \sim c$. This confirms the argument mentioned in Section \ref{sec:intro} that the acceleration of electrons via the DSA with MHD turbulence requires a seed population of mildly relativistic energies in typical interplanetary and interstellar plasma conditions. The scattering of non-relativistic electrons $v \ll c$ needs shorter wavelength modes. For the right-hand polarized mode, the resonance occurs in the wavenumber range of whistler-mode waves $k \lambda_{\rm i} \gtrsim 1$.

In the following, $k_{\rm res} = k_{\rm res} (v, \mu)$ denotes the resonant wavenumber for given energy and pitch angle determined by Equations (\ref{resonance}) and (\ref{dwhistler}). The standard QLT relates $D_{\mu\mu}$ to the power spectrum of waves as follows \citep[e.g.][]{1972ApJ...172..319J,1975MNRAS.172..557S}:
\begin{align} \label{power}
 D_{\mu\mu}(v,\mu)=\frac{\pi}{4}\frac{I(k_{{\rm res}})k_{{\rm res}}}{B_{\rm 0}^2}\Omega_{{\rm ce}}.
\end{align}
Note that $I(k_{\rm res})dk/2\mu_{\rm 0}$ is the energy density of magnetic field fluctuations in interval $dk$ evaluated at $k = k_{\rm res}$, where $\mu_0$ represents the permeability of free space. This describes energy and pitch-angle dependence of $D_{\mu\mu} (v,\mu)$ through $k_{\rm res} (v,\mu)$.

We have estimated the cut-off energy of the accelerated electrons by Equation (\ref{cutoff}) under the assumption of constant $D_{\mu\mu}$. Looking at the condition from a different perspective, we may obtain the inequality
\begin{align}\label{rcutoff}
 \frac{D_{\mu\mu}}{\Omega_{\rm ce}} \gtrsim
 \left( \frac{m_{\rm e}}{m_{\rm i}} \right)
 \left( \frac{\epsilon}{\epsilon_{\rm sh}} \right),
\end{align}
which must be satisfied for the acceleration process to continue at this particular particle energy $\epsilon$. In reality, this condition must be satisfied in the whole energy range in between the injection $\epsilon_{\rm inj}$ and the maximum cut-off $\epsilon_{\rm max}$ where the spectrum forms a power-law. Equivalently, we obtain the threshold wave power by using the QLT estimate of $D_{\mu\mu}$ (Equation (\ref{power}))
\begin{align} \label{pcutoff}
 \frac{I(k_{{\rm res}})k_{{\rm res}}}{B_0^2}
 \gtrsim
 \frac{4}{\pi}
 \left( \frac{m_{\rm e}}{m_{\rm i}} \right)
 \left( \frac{\epsilon}{\epsilon_{\rm sh}} \right),
\end{align}
Noting that the resonant wavenumber $k_{\rm res}$ is a monotonically decreasing function of energy $\epsilon$ (for a fixed $\mu$), we see that the wave power must always be larger than the threshold in the range $k_{\rm res} (\epsilon_{\rm max}) < k < k_{\rm res} (\epsilon_{\rm inj})$ in order to sustain the power-law spectrum in the corresponding energy range.

\begin{figure}[ht!]
\centering
\includegraphics[width=0.45\textwidth]{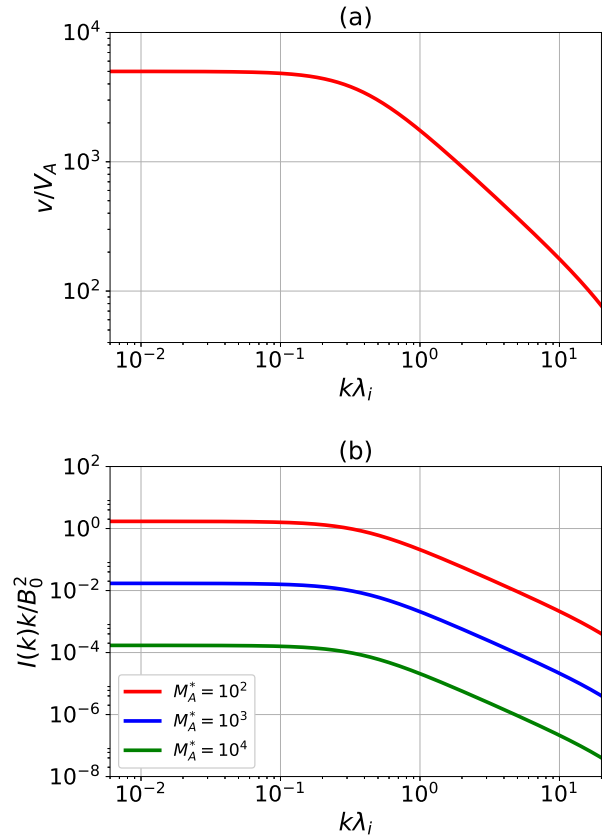}
\caption{(a) The relationship between the resonance velocity and wavenumber obtained with $V_A/c=2\times10^{-4}$. (b) The threshold wave power which is required to sustain the SSDA as a function of $k$. The results obtained with $M_{\rm A}^*=10^2,\,10^3,\,10^4$ are shown.}
\label{fig:tpower}
\end{figure}

The dependence of the maximum energy on $\epsilon_{\rm sh} \propto u_{\rm sh}^2$ and the resonant wavenumber $k_{\rm res}$ on $V_{\rm A}/c$ suggest that it is natural to introduce an effective Mach number $M_{\rm A}^{*} \equiv u_{\rm sh}/V_{\rm A} = M_{\rm A}/\cos \theta_{B_{\rm n}}$. For a fixed value of $V_{\rm A}/c$, $M_{\rm A}^{*}$ is the only macroscopic shock parameter that regulates the problem. Figure \ref{fig:tpower}(b) shows the threshold wave power as a function of wavenumber for selected values of $M_{\rm A}^{*}=10^2,10^3,10^4$ obtained with a fixed $V_{\rm A}/c=2\times10^{-4}$. Figure \ref{fig:tpower}(a) shows the resonance energy as a function of wavenumber. It is clear that the threshold to sustain the acceleration at a normalized energy $\epsilon/\epsilon_{\rm sh}$ decreases as increasing the effective Mach number $M_{\rm A}^{*}$. In other words, for a given strength of wave power, the range of applicability of the SSDA model will extend as increasing $M_{\rm A}^{*}$. Therefore, quasi-perpendicular and high Mach number shocks are favored not only from the viewpoint of the classical SDA but also from the effectiveness of scattering.

From Figure \ref{fig:tpower}(b), we see the wavenumber dependence of the threshold $I(k) \propto k^{-2}$ ($I(k) k \propto k^{-1}$) in the whistler-mode range $k \lambda_i \gtrsim 1$. If we assume a power-law of the wave power spectrum which is steeper than $\propto k^{-2}$, the condition is the most stringent at large wavenumber (or low energy). If the condition is satisfied at some wavenumber $k^{*}$, it continues to be so at smaller $k < k^{*}$. In this case, once the particles obtain the threshold energy determined by $k^{*}$, the acceleration will continue until $D_{\mu\mu}$ saturates by, e.g., nonlinearity ($\delta B \sim B_0$). In contrast, a shallower spectrum implies that the maximum energy is limited by the threshold wave power at a low wavenumber, whereas there is no lower bound in energy. Although it is virtually impossible to estimate the spectrum theoretically within the shock transition region, one might presume that the spectrum may be steeper than the threshold because of high wavenumbers (where dissipation is significant). In any case, it is true that an efficient acceleration from the thermal energy requires high-frequency whistler waves of sufficiently large amplitudes.

\begin{figure}[ht!]
\centering
\includegraphics[width=0.45\textwidth]{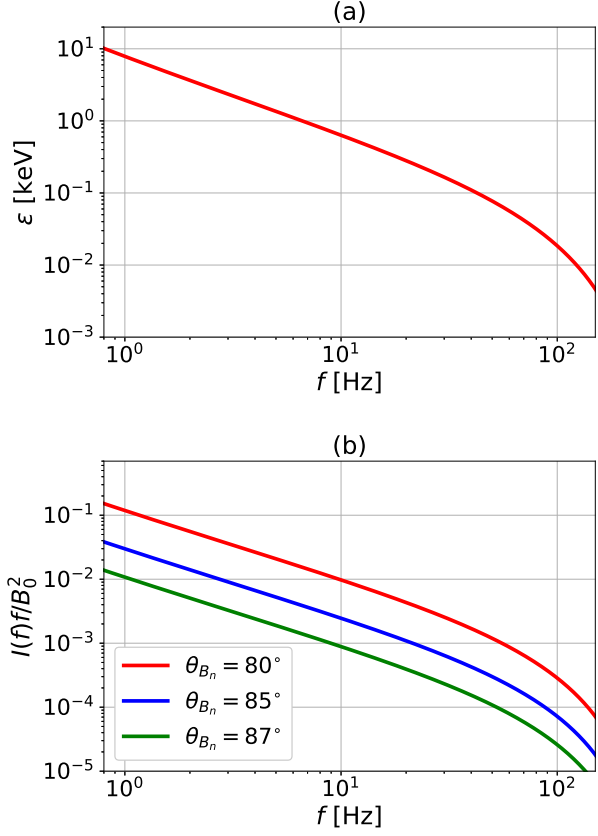}
\caption{(a) The relationship between the resonant energy and wave frequency obtained with $B=10\,\text{nT}$, $n_{e}=10\,\text{cm}^{-3}$, and $u_0=400\,\text{km/s}$ (typical parameters for Earth's bow shock). (b) The threshold wave power as a function of wave frequency for three different values of $\theta_{B_{n}}=80\degr, 85\degr, 87\degr$.}
\label{fig:bowshock}
\end{figure}

We now consider the application of the theory to Earth's bow shock. Figure \ref{fig:bowshock} shows the threshold wave power as a function of frequency $f = \omega/2\pi$ and the relation between the frequency and the resonance particle energy calculated using typical parameters of the bow shock: an upstream magnetic field strength of $B_0 = 10 \, {\rm nT}$, density of $n_{\rm e} = 10 \, {\rm cm^{-3}}$, and solar wind speed of $u_{\rm 0} = 400 \, {\rm km/s}$. Note that we use the frequency in the plasma rest frame which is, in general, different from that measured in the spacecraft frame. The relative contribution of the Doppler shift in frequency may be estimated as $\delta f/f \sim k u_0 \cos \theta_{\rm kb}/\omega$ where $\theta_{\rm kb}$ is the angle between the magnetic field and the wave propagation directions. For high-frequency whistler waves propagating along the magnetic field, this effect is not significant. However, this may not be true for lower frequency obliquely propagating modes.

We can see from Figure \ref{fig:bowshock}(a) that the resonant frequency for electrons with $\epsilon \sim 0.1 \, {\rm keV}$ and $\sim 10 \, {\rm keV}$ (energy range of our primary interest) are $f \sim 40 \, {\rm Hz}$ and $\sim 1 \, {\rm Hz}$, respectively. In reality, since the magnetic field is compressed a few times within the shock transition region compared to the upstream ($B = 10 \, {\rm nT}$ in this example), the corresponding frequencies are also higher by the same factor. Therefore, the typical frequency of whistlers that are needed to pitch-angle scatter low-energy electrons $\epsilon \lesssim 1 \, {\rm keV}$ is roughly $f \sim 100 \, {\rm Hz}$. In the following discussion, we focus only on the high-frequency band because observationally we know that the power at $\sim 1 {\rm -} 10 \, {\rm Hz}$ (around lower-hybrid frequency) is substantial ($\delta B \sim B_{\rm 0}$) and the threshold at the higher frequency imposes a more stringent condition for the particle acceleration.

Figure \ref{fig:bowshock}(b) shows that the wave power required at the high-frequency band is around $I(f) f / B_0^2 \sim 10^{-4} {\rm -} 10^{-2}$ for $\theta_{B_{\rm n}} \gtrsim 80 \degr$. This may be compared with in-situ detection of high-frequency whistlers observed within quasi-perpendicular bow shocks \citep{1999JGR...104..449Z,2012JGRA..11712104H,2017ApJ...842L..11O}. These observations found intense and relatively narrowband whistlers at frequencies $f \gtrsim 100 \, {\rm Hz}$ that were nearly parallel propagating along the ambient magnetic field. Characteristic wave amplitudes and frequencies were $\delta B/B_{\rm 0} \sim 0.01 {\rm -} 0.1$ and $\omega/\Omega_{ce} \sim 0.1 {\rm -} 0.4$ in normalized units \citep{1999JGR...104..449Z}. Assuming narrowband waves with the frequency bandwidth $\Delta f$ satisfying $\Delta f/f \lesssim 1$, one may approximate $I(f) \sim \delta B^2 / \Delta f$. We thus obtain
\begin{align}
 \frac{I(f) f}{B_{\rm 0}^2} \sim
 \left( \frac{\delta B}{B_{\rm 0}} \right)^2
 \left( \frac{f}{\Delta f} \right)
 \sim 10^{-4} {\rm -} 10^{-2},
\end{align}
which indicates that intense whistlers at high frequencies provide sufficient scatterings for accelerating low-energy electrons $\epsilon \lesssim 1 \, {\rm keV}$ only at nearly perpendicular shocks $\theta_{B_{\rm n}} \gtrsim 80 \degr$.

If the high-frequency whistlers have sufficient amplitudes, the power at lower frequencies is in general much larger. Therefore, the SSDA will continue to accelerate electrons to higher and higher energies. The scattering efficiency due to large-amplitude fluctuations $\delta B \sim B_{\rm 0}$ at low frequencies may in principle reach the Bohm scattering limit $D_{\mu\mu}/\Omega_{\rm ce} \sim 1$, which is, however, too optimistic. Instead, we take a more realistic value of the maximum pitch-angle scattering $D_{\mu\mu}/\Omega_{\rm ce} \sim 0.1$ to estimate the maximum energy. Using typical parameters of the bow shock, we obtain the maximum energy:
\begin{align}
 \epsilon_{\rm max} \approx 11 \, {\rm keV} \,
 \left( \frac{u_{\rm 0}}{400 \, {\rm km/s}} \right)^2
 \left( \frac{\cos \theta_{B_{\rm n}}}{\cos 85 \degr} \right)^{-2}
 \left( \frac{D_{\mu\mu}}{0.1\,\Omega_{\rm ce}}\right)
\end{align}
This is consistent with observations at the bow shock, in which the power-law spectra of the accelerated electron continue to $\gtrsim 10 \, {\rm keV}$ at nearly perpendicular shocks.

Note that we have assumed a plane shock in the foregoing discussion. It is known that a finite curvature of the bow shock may limit the maximum energy of the SDA. This is because the accelerated electrons travel a long distance along the magnetic field line, and eventually they will escape from the shock due to the change in local $\theta_{B_{\rm n}}$. The same argument can apply for the SSDA, which will limit the maximum energy of electrons to $\sim 10 \, {\rm keV}$ at the bow shock \citep{1991JGR....96..143K}. This is obviously not an intrinsic limit of the SSDA mechanism, and shocks with larger scale sizes will be able to accelerate electrons to higher energies.

Given the scaling law of the maximum energy as a function of shock speed, we believe that the SSDA mechanism may accelerate electrons to $\sim 1 \, {\rm MeV}$ if the shock speed is ten times higher than the bow shock. Young SNR shocks can naturally satisfy the condition. Therefore, we think that the electron acceleration in such high-speed shocks may be initiated by the SSDA in the shock transition region up to mildly relativistic energies. The relativistic electrons escaped from the shock transition region may be further accelerated by the subsequent DSA, producing highly relativistic electrons that are seen in synchrotron emission. The SSDA model thus provides a plausible solution to the electron injection problem.

As we have seen, the energy and pitch-angle dependence of the pitch-angle scattering coefficient is of critical importance for the injection. Although we have referred to the wave amplitudes reported previously to estimate the pitch-angle scattering efficiency, we need to understand the wave generation mechanisms for better predictability.

One of the possible mechanisms for the generation of whistler waves is the modified-two-stream instability (MTSI) driven by the reflected ion beam streaming across the ambient magnetic field. Full-particle simulations \citep{2003JGRA..108.1459M,2006JGRA..111.6104M} have shown that the MTSI can generate low-frequency ($\lesssim 0.05\Omega_{\rm ce}$) whistler waves propagating obliquely with respect to the magnetic field. Therefore, the MTSI is a plausible candidate for generating waves with frequencies below $\sim$ 10 Hz at Earth's bow shock. However, such low-frequency waves can scatter only electrons with energy $\gtrsim$ 10 keV. The scattering of lower energy electrons
need higher frequency whistlers ($\gtrsim0.1\Omega_{\rm ce}$).

On the other hand, the instability driven by a finite electron temperature anisotropy may also generate high-frequency whistler waves. Because of the conservation of the first adiabatic invariant in the magnetic field compression at the shock, electrons are adiabatically heated primarily in the direction perpendicular to the magnetic field. This tends to produce a higher perpendicular temperature than a parallel temperature. Such a perpendicular temperature anisotropy may destabilize high-frequency whistlers via the electron cyclotron resonance. In reality, the electron distribution function in the shock transition layer may be much more complicated than the simple temperature anisotropy. It may have a beam and a loss-cone, both of which affect the stability of whistler waves \citep[e.g.,][]{1984JGR....89..105T,2010PhRvL.104r1102A}. These details must be taken into account to understand the generation mechanism for high-frequency ($\sim$ 100 Hz) coherent whistler waves that are the only agent for scattering low-energy ($\lesssim$ 1 keV) electrons.

\section{Summary} \label{sec:summary}
We have proposed the stochastic shock drift acceleration (SSDA) model that may explain the acceleration of sub-relativistic electrons observed within the transition layer of planetary bow shocks. The process naturally predicts a power-law energy spectrum with a maximum energy cut-off. The spectral index does not depend on the pitch-angle scattering coefficient $D_{\mu\mu}$, whereas the maximum energy is linearly proportional to $D_{\mu\mu}$. Monte-Carlo simulations have confirmed the theoretical predictions.

By referring to literature on in-situ observations of waves in the shock transition region, we have estimated $D_{\mu\mu}$ using the standard quasi-linear theory (QLT). The result indicates that whistler waves observed within Earth's bow shock have sufficiently large amplitudes to provide the necessary pitch-angle scattering when the shock is nearly perpendicular $\theta_{B_{\rm n}} \gtrsim 80 \degr$.

The maximum energy at the bow shock is on the order of a few tens of keV for typical solar wind conditions. The theoretical scaling law suggests that SNR shocks may accelerate electrons up to relativistic energies. Therefore, the electron injection to subsequent diffusive shock acceleration (DSA) may be achieved through the SSDA at such high-speed shocks. More detailed analyses using first-principles PIC simulations and in-situ measurements need to be carried out to confirm the applicability of this model.
\acknowledgements
The authors are grateful to M. Hoshino, Y. Matsumoto, and Y. Ohira for fruitful comments and discussions. This work was supported by JSPS KAKENHI Grant Numbers 17H02966, 17H06140.
\appendix
\section{Derivation of Equation (5)} \label{sec:appendix}
We estimate the jump in the magnetic field strength and fluid velocity parallel to the magnetic field at the shock using the Rankine-Hugoniot relation \citep[e.g.,][]{gurnett_bhattacharjee_2005} as follows: 
\begin{align}
\frac{\Delta B}{B_{\rm 0}}&=\sqrt{r\left(1+\frac{(r-1)(M_{\rm A}^2\cos^2{\theta_{B_{\rm n}}}+r)}{M_{\rm A}^2\cos^2{\theta_{B_{\rm n}}}-r}\right)}-1\notag\\
&\approx r-1, \label{b_app}\\
\frac{\Delta u_{\parallel}}{u_{\rm sh}}&=1-\sqrt{\left(1+\frac{r-1}{M_{\rm A}^2\cos^2{\theta_{B_{\rm n}}}-r}\right)^2\sin^2{\theta_{B_{\rm n}}}+\frac{1}{r^2}\cos^2{\theta_{B_{\rm n}}}}\notag\\
&\approx1-\sqrt{1-\left(1-\frac{1}{r^2}\right)\cos^2{\theta_{B_{\rm n}}}} \notag\\
&\approx \frac{1}{2}\left(1-\frac{1}{r^2}\right)\cos^2{\theta_{B_{\rm n}}}+O(\cos^4{\theta_{B_{\rm n}}}), \label{v_app}
\end{align}
where $\Delta B,\Delta u_{\parallel}$ represent the absolute values of the jump in $B,u_{\parallel}$ across the shock, and $r$ is the shock compression ratio. Note that we assumed a high Mach number ($M_{\rm A}\gg1$) and quasi-perpendicular shock ($\cos{\theta}\ll 1$). We see that the magnetic field jump is on the order unity, whereas the parallel velocity jump is small, on the order of $\cos^2 \theta_{B_{\rm n}}$.

By using these estimates and the length scale of the shock transition region $L_{\rm sh}$, the gradients of $\log{B}$ and $\log{u_{\parallel}}$ can be approximated by
\begin{align}
\frac{\partial \log B}{\partial s}&\approx\frac{1}{L_{\rm sh}}\log{\left(1+\frac{\Delta B}{B_{\rm 0}}\right)} \label{es1}\notag\\
&\approx\frac{\log{r}}{L_{\rm sh}},\\
\frac{\partial \log{u_{\parallel}}}{\partial s}&\approx\frac{1}{L_{\rm sh}}\log{\left(1-\frac{\Delta u_{\parallel}}{u_{\rm sh}}\right)}\label{es2}\notag\\
 &\approx\frac{1}{L_{\rm sh}}\log{\left[1-\frac{1}{2}\left(1-\frac{1}{r^2}\right)\cos^2{\theta_{B_{\rm n}}}+O\left(\cos{\theta_{B_{\rm n}}^4}\right)\right]}\notag\\
&=-\frac{1}{2L_{\rm sh}}\left(1-\frac{1}{r^2}\right)\cos^2{\theta_{B_{\rm n}}}+O\left(\cos^4{\theta_{B_{\rm n}}}\right).
\end{align}
From these, we finally obtain Equation (\ref{estimation}) 
\begin{align}
 \left|\left(\frac{\partial \log{u_{\parallel}}}{\partial s}\right)\left(\frac{\partial\log{B}}{\partial s}\right)^{-1}\right|&\approx\frac{1}{2\log{r}}\left(1-\frac{1}{r^2}\right)\cos^2{\theta_{B_{\rm n}}}+O\left(\cos^4{\theta_{B_{\rm n}}}\right).
\end{align}

\bibliography{reference}

\end{document}